\documentstyle[iopconf1]{article}
\begin{document}
\title{Conformality and Brodsky}

\author{Paul H. Frampton 
\footnote{E-mail:
frampton@physics.unc.edu}}

\affil{Department of Physics and Astronomy,
University of North Carolina,
Chapel Hill, NC 27599-3255, USA.}

\beginabstract
In this article I describe the recently-conjectured
field-string duality which 
suggests a class of nonsupersymmetric gauge theories
which are conformal (CGT) to leading order of 1/N and some of which
may be conformal for finite N. 
If the standard model becomes conformal at
TeV scales, model-building
on this basis is an interesting direction. Some remarks
are added about the inspiring career of Stan Brodsky.
\endabstract

\section{Introductory remark}

It is irresistible to take advantage of a typographical error
in the printed program for this meeting where my title
was off by two letters:
``Conformity and Brodsky". Of course I would be very pleased if in ten years
time at Brodsky's 70th birthday conformality had
become conformity! As will become clear below conformality is
antithetical to the presently most popular
ideas for extending the standard model, {\it viz.} 
grand unification and lessens
the motivation for another popular idea,
{\it viz.} low-energy supersymmetry.

\section{Abbreviated History of String Theory}

The recent
development of field string duality
possesses some quality of {\it d\'{e}ja vu} and yet seems
the most promising development in the theory in terms
of its most optimistic prognosis that it may provide a successful
connection between string theory and the real world, and in
doing so necessarily a first connection between gravity
and the other interactions.

The initial seed of string theory was the Veneziano model
in 1968\cite{veneziano}. At the time, finite energy  and
superconvergence sum rules for hadron scattering (the 
subject of my DPhil thesis) posed a "duality"
of descriptions generally similar to the now-proposed one
between a ten-dimensional superstring (or 11 dimensional M theory)
and a conformal gauge theory. The hadron sum rules equated quantities
of quite different functional dependences on the Mandelstam variables
$s$ and $t$, seemingly an impossibility until the 
Veneziano model showed an explicit realization.

From 1968 to 1973 the resultant dual resonance models
were leading candidates to describe the strong interactions.
In 1973 they were, however, elbowed aside by an alternative  
theory, quantum chromodynamics (QCD). 
Now the discarded theory
is dual to the QCD which replaced it!\cite{maldacena,ooguri}. This is what
I mean by {\it d\'{e}ja vu}. 

The decade 1974-1984 saw a hiatus in string theory.
In 1984-85 the First Superstring Revolution included a stab at 
nearly a Theory of Everything, the perturbative $E(8)\times E(8)$
heterotic string\cite{candelas}. But its  apparent uniqueness turned
out to be illusory (as was its perturbativity), and another decade
1985-95 of quiescence followed.

In 1995 came the Second Superstring evolution, and in 1997 the
$2\frac{1}{2}$-revolution with AdS/CFT duality. 
Understanding of duality between weak and strong coupling of
supersymmetric field theories led to a corresponding breakthrough in
string theory culminating with the idea of M theory as a more basic theory
which unified all of the five known ten-dimensional superstings
(Types I, IIA, IIB and the $O_{32}$ and $E_8\times E_8$
heterotic strings) as well as eleven-dimensional supergravity
by duality transformations.

One of the most important realizations of the recent period is
that string solitons, or D branes, play a dynamical role in
the theory equally as important 
as do the superstrings themselves. D branes
are crucial for the field-string duality which is our principal 
subject. The string duality has also led to a better understanding
of the quantum mechanics of black holes.

Our starting point here is the duality
between string theory in 10 dimensions (or of M theory in 11 dimensions)
and gauge field theory in four dimensions. As already mentioned,
this in a real sense closes a 25-year cycle in the history of
strings.

Certainly one of the major changes in string theory in the recent years
is the appreciation of the role of D branes which are topological defects
on which open strings can end. Their necessity in string theory
was realized only in 1989\cite{polchinski} and particularly
in 1995\cite{polchinski2}. Their presence follows from considering
the $R \rightarrow 0$ limit of a bosonic string compactified
on a circle of radius $R$. Open strings, unlike the closed strings
which are necessarily contained in the same theory, cannot wrap
around the compactified dimension in the $R \rightarrow 0$ limit.
Hence for a consistent theory the open strings do not simply end, but
are attached to D branes.

These D branes have their own dynamics and play a central role in
the full non-perturbative theory. D branes have provided insight into
(1) Black hole quantum mechanics; (2) Large N gauge field theory
(discussed here).

In general, the term duality applies to a situation where two quite
different descriptions are available for the same physics.

The difference can be very striking. For example, in 1997
Maldacena\cite{maldacena} proposed the duality between
$d = 4$ $SU(N)$ gauge field theory (GFT) and a $d = 10$ superstring.
In the perturbative regime of the GFT this duality cannot hold just because the
degrees of freedom are missing, but non-perturbatively the GFT
contains sufficient additional states at strong coupling for the
duality to be indeed possible.

Take a Type IIB superstring (closed, chiral) in $d = 10$ and compactify
it on the manifold:
\begin{equation}
(AdS)_5 \times S^5
\end{equation}
Here $(AdS)_5$ is a 5-dimensional Anti De Sitter space whose four dimensional
surface $M_4$ is the $d = 4$ spacetime in which the $SU(N)$ GFT occurs. 
Note that the isometry group of $(AdS)_5$ is $SO(4,2)$, the conformal group
for four dimensional spacetime. The $S^5$ is a five-sphere with isometry $SU(4)$ which is the R symmetry
of the resultant ${\cal N} = 4$ supersymmetric $SU(N)$ GFT. The
$S^5$ can be regarded as a surface in a $C_3$ three-dimensional space
in which $N$ D3 branes are coincident.

The D branes each carry an associated $U(1)$ gauge symmetry. This
is understandable as a correct generalization 
of the Chan-Paton factors\cite{chan}
which were once used to attach charges to the ends of open strings.
$N$ parallel D branes with vanishing separation yield a $U(N)$
gauge group where the additional $N^2 - N$ gauge bosons arise from connecting
open strings which become massless in the zero-length limit. This
$U(N)$ turns out to be broken to $SU(N)$ by the brane dynamics.
The resultant ${\cal N} = 4$ SUSY Yang-Mills theory is well-known to
be a very well-behaved, finite field theory. It is conformally
invariant even for finite $N$ with all RG $\beta-$ functions 
(gauge, Yukawa and quartic Higgs) vanishing.

This perturbative finiteness was proved in 1983 by Mandelstam\cite{mandelstam}.
The Maldacena conjecture is primarily aimed at the $N \rightarrow \infty$
limit with the 't Hooft parameter\cite{hooft} of $N$ times the squared gauge
coupling fixed, and makes no claim concerning conformality
for finite $N$\cite{maldacena2}. But since the ${\cal N} = 4$ case
is known to be conformal even for finite $N$ one is tempted
to extend the conjecture\cite {frampton} to finite $N$ cases even where
all supersymmetry is broken. In that case the standard model can be a part
of a conformal nonsupersymmetric gauge theory where the
$\beta-$ functions become zero at a TeV scale. Then the coupling 
constants cease to run and there is no 
grand unification. This nullifies the gauge
hierarchy problem between the weak scale and the GUT scale,
and yet it is still possible to derive the correct electroweak
mixing angle\cite{frampton4}. In particular there is no reason to invoke
low-energy supersymmetry either. Gravity is itself non-conformal
(it necessitates the dimensionful Newton constant). We shall
address this at the end of the article.

\section{Breaking Supersymmetries.}

To approach the real world one needs less supersymmetry than
${\cal N} = 4$, in fact the empirical data presently suggest no
supersymmetry at all, ${\cal N} = 0$.

By factoring out a discrete group (we shall
assume it is an abelian discrete group, only
because that case has been most fully investigated; it
is possible that a non-abelian discrete group can work as well)
and composing the orbifold:
\begin{equation}
S^5/\Gamma
\end{equation}
one may break ${\cal N} = 4$ supersymmetry to
${\cal N} = 2, 1 ~ {\rm or} ~ 0$. Of special interest
is the ${\cal N} = 0$ case.

We may take $\Gamma = Z_p$ which identifies $p$ points in $C_3$.

The rule for breaking the ${\cal N}=4$
supersymmetry is:
\begin{equation}
\Gamma \subset SU(2) \Longrightarrow {\cal N} = 2
\end{equation}
\begin{equation}
\Gamma \subset SU(3) \Longrightarrow {\cal N} = 1 
\end{equation}
\begin{equation}
\Gamma \not\subset SU(3) \Longrightarrow {\cal N} = 0 
\end{equation}

In fact, to specify the embedding of $\Gamma = Z_p$, we
need to identify three integers $a_i = (a_1, a_2, a_3)$
such that the action of $Z_p$ on $C_3$ is:
\begin{equation}
C_3:  (X_1, X_2, X_3) \stackrel{Z_p}{\longrightarrow} 
(\alpha^{a_1} X_1, \alpha^{a_2} X_2, \alpha^{a_3} X_3 ) 
\end{equation}
with
\begin{equation}
\alpha = exp \left( \frac{2 \pi i}{p} \right)
\end{equation}

The scalar multiplet is in the {\bf 6} of $SU(4)$
R symmetry and is transformed by the $Z_p$ transformation:
\begin{equation}
diag (\alpha^{a_1}, \alpha^{a_2}, \alpha^{a_3}, \alpha^{-a_1}, \alpha^{-a_2}, \alpha^{-a_3})
\end{equation}
together with the gauge transformation
\begin{equation}
diag (\alpha^0, \alpha^1, \alpha^2, \alpha^3, \alpha^4, \alpha^5) ~~~\times ~~~ \alpha^i
\end{equation}
for the different $SU(N)_i$ of the gauge group $SU(N)^p$.

What will be relevant are states invariant under a combination of
these two transformations, as discussed in the next subsection.

If $a_1 + a_2 + a_3 = 0 ~~~ ({\rm mod ~~~ p})$ then the matrix
\begin{equation}
\left( \begin{array}{ccc}
a_1 & & \\
 & a_2 &  \\
  &  &  a_3 
\end{array} \right)
\end{equation}
is in $SU(3)$ and hence ${\cal N} \geq 1$ is unbroken and this
condition must therefore be
avoided if we want ${\cal N} = 0$.

If we examine the {\bf 4} of $SU(4)$, we find that the matter
which is invariant under the combination of the $Z_p$ and an
$SU(N)^p$ gauge transformation can be deduced similarly.

It is worth defining the spinor {\bf 4} explicitly by
$A_q = (A_1, A_2, A_3, A_4)$ with the $A_q$, like the $a_i$, defined only
${\rm ~ mod ~ p}$. Explicitly we may define
$a_1 = A_1 + A_2, ~~ a_2 = A_2 + A_3, ~~ a_3 = A_3 + A_1$ and
$A_4 = - (A_1 + A_2 + A_3)$. In other words,
$A_1 = \frac{1}{2}(a_1 - a_2 + a_3)$,
$A_2 = \frac{1}{2}(a_1 + a_2 - a_3)$,
$A_3 = \frac{1}{2}(-a_1 + a_2 + a_3)$,
and $A_4 = - \frac{1}{2}(a_1 + a_2 + a_3)$. To leave no unbroken
supersymmetry we must obviously
require that all $A_q$ are non-vanishing. In terms of the $a_i$ this
condition which we shall impose is:
\begin{equation}
\sum_{i = 1}^{i = 3} \pm (a_i) \neq 0 ~~~({\rm mod ~~~ p})
\label{modp}
\end{equation}

The question at issue is whether the gauge theories derived in
this way are conformal for finite N. What is known is that at
leading order in $1/N$ the $\beta-$ functions vanish to all
orders in perturbation theory\cite{BJ}. This is already remarkable
from the field theory point of view because without the stimulus of
the AdS/CFT duality it would be difficult to guess any 
${\cal N} = 0$ theory with all $\beta-$ functions zero
to leading order in $1/N$ and all orders in the GFT coupling.
Without non-renormalization theorems this imposes an
infinite number of constraints on a finite number of
choices of the fermion and scalar representations
of $SU(N)^p$.

Nevertheless, since ${\cal N} = 4$ is conformal (all $\beta-$ functions
vanish) we can be more ambitious and ask\cite{frampton} that
all $\beta-$ functions vanish even for finite $N$, at least for
some fixed point in coupling constant space, and use the construction
to motivate phenomenological model-building.

\section{Matter Representations.}

The $Z_p$ group identifies $p$ points in $C_3$. The $N$ converging
D branes approach all $p$ such points giving a gauge
group with $p$ factors:
\begin{equation}
SU(N) \times SU(N) \times SU(N) \times  . . . . \times SU(N) 
\end{equation}
The matter which survives is invariant under a product of a gauge transformation
and a $Z_p$ transformation.

For the covering gauge group $SU(pN)$, the transformation is:
\begin{equation}
(1, 1, ... ,1; \alpha, \alpha, .... \alpha;
\alpha^2, \alpha^2,  ...  \alpha^2; . . . . . . ;
\alpha^{p - 1}, \alpha^{p - 1}, . . . \alpha^{p - 1})
\end{equation}
with each entry occurring $N$ times.

Under the $Z_p$ transformation for the scalar fields, the {\bf 6}
of $SU(4)$, the tranformation is
\begin{equation}
 \sim~~\underline{X} \Rightarrow ~~ (\alpha^{a_1}, \alpha^{a_2},
\alpha^{a_3})
\end{equation}

The result can conveniently be summarized
by a {\it quiver diagram}\cite{DM}. One draws $p$ points
and for each $a_k$ one draws a non-directed arrow between
all modes $i$ and $i + a_k$. Each arrow denotes a bi-fundamental representation
such that the resultant scalar representation is:
\begin{equation}
\sum_{k=1}^{k=3} \sum_{i=1}^{i=p} (N_i, \bar{N}_{i \pm a_k})
\end{equation}
If $a_k=0$ the bifundamental is to be reinterpreted as an adjoint
representation plus a singlet representation.

For the chiral fermions one must construct the spinor {\bf 4}
of $SU(4)$. The components are the $A_q$ given above. The resultant
fermion representation follows from a different quiver diagram.
One draws $p$ points and connects with a {\it directed} arrow
the node $i$ to the node $i + A_q$. The fermion represntation
is then:
\begin{equation}
\sum_{q=1}^{q =4} \sum_{i=1}^{i=p} (N_i, \bar{N}_{i+A_q})
\end{equation}
Since all $A_q \neq 0$, there are no adjoint
representations for fermions.
This completes the matter representation of $SU(N)^p$.

\bigskip

We have begun the selection process between these models
by looking at one
and two loop $\beta-$functions. At one loop we are still at leading order in $N$
at least for $\beta_g$ so there is coincidence with
the ${\cal N} =4$ case. At 2 loops we found\cite{FS} already that only
$8\%$ of a sample satisfy one criterion, the fraction remaining alive diminishing
like $3/p$ for large $p$.

Checking the Yukawa and Higgs running for 2 loops
needs more calculation of couplings and is underway.

Beyond that:
\begin{itemize}

\item If all 2-loop tests are satisfied, what about 3 or more?
It rapidly becomes impractical to take the approach of direct calculation.

\item There is the question of uniqueness of any surviving
${\cal N} = 0$ CGT.

\item The CGT may be inspirational in model building, to be discussed below.

\end{itemize}

{\it Why ${\cal N} = 0$ ?}.

${\cal N} = 1$ is motivated by accommodation of the gauge
hierarchy $M_{GUT}/M_{WEAK}$.

In a conformal gauge theory the gauge couplings cease to run
and the GUT scale does not exist; this hierarchy is therefore nullified.

Low-scale Kaluza-Klein\cite{antoniadis} is similar to the conformality approach
in this particular regard, although the idea is quite different.

More philosophically, we may recall the over 50 years ago
the infinite renormalization of QED was greeted with much
skepticism. If the conformality of even ${\cal N} = 4$
CFT had been already discovered, surely the skepticism would
have been far greater?

\bigskip
\bigskip

\section{Conformality and Particle Phenomenology.}

Let us itemize the following points:

\begin{itemize}

\item The hierarchy between the GUT and weak scales is 
14 orders of magnitude.

\item Why do the two very different scales exist?

\item How are the scales stabilized under quantum corrections?

\item Supersymmetry solves the second problem but not the first.

\end{itemize}

{\it Successes of supersymmetry.}

\begin{itemize}

\item Cancellations of UV infinities.

\item Technical naturalness of hierarchy.

\item Unification of gauge couplings.

\item Natural appearance in string theory.

\end{itemize}

{\it Puzzles about supersymmetry.}

\begin{itemize}

\item The ``mu" problem - why is the Higgs mass at the weak scale and
not at the Planck scale (hierarchy).

\item Breaking supersymmetry leads to too large a cosmological constant.

\item Is supersymmetry fundamental to string theory?

\item There are solutions of string theory without supersymmetry.

\end{itemize}

{\it Supersymmetry replaced by conformality at TeV scale.}

The following aspects of the idea are
discussed:

\begin{itemize}

\item The idea is possible.

\item Explicit examples containing standard model states.
\item Finiteness as a more rigid constraint than supersymmetry.

\item Predicts additional states for finiteness/conformality.

\item Rich inter-family structure of Yukawa couplings.

\end{itemize}

\section{Conformality as Hierarchy Solution.}

The quark and lepton masses, the QCD scale and the weak scale are 
extremely small compared to a TeV scale. 
They may all be put to zero suggesting: add
degrees of freedom to yield GFT with conformal invariance 
(CGT). 't Hooft's
naturalness condition holds since zero mass 
increases the symmetry.

The theory is assumed to be given by the action\cite{FV}
\begin{equation}
S = S_0 + \int d^4x \alpha_i O_i
\end{equation}
where $S_0$ is the action for the conformal theory and the $O_i$ are operators with
dimension below four which break conformal invariance softly.

The mass parameters $\alpha_i$ have mass dimension 
$4 - \Delta_i$ where $\Delta_i$ is the dimension of $O_i$ at the conformal point.

Let $M$ be the scale set by the parameters $\alpha_i$ and hence the scale at which conformal invariance
is broken. Then for $E \gg M$
the couplings will not run while they start
running for $E < M$. To solve the hierarchy problem we assume $M$ is 
close to the TeV scale.

\section{Large Class of d=4 QFTs - Each SU(4) Subgroup.}

There is first the choice of $N$. 
One knows that for leading $1/N$ the theory is conformal.
What about finite N? One expects at least a conformal 
fixed point in some cases.
One starts from ${\cal N} = 4$ GFT, eliminates fields and re-identifies others
such that conformality results.

It is important to realize that, even {\it without} supersymmetry,
boson-fermion number equality holds, and underlies the finiteness.

Let $\Gamma\subset SU(4)$ denote a discrete subgroup of $SU(4)$.
Consider irreducible representations of $\Gamma$.  Suppose there
are $k$ irreducible representations $R_i$, with dimensions $d_i$ with
$i=1,...,k$.  The gauge theory in question has gauge symmetry
$$SU(N d_1)\times SU(Nd_2)\times ...SU(Nd_k)$$
The fermions in the theory are given as follows.  Consider the 4 dimensional
representation of $\Gamma$ induced from its embedding in $SU(4)$.  It
may or may not be an irreducible representation of $\Gamma$. We consider
the tensor product of ${\bf 4}$ with the representations $R_i$:

\[
{\bf 4}\otimes R_i = \oplus_j n_i^jR_j
\]

The chiral fermions are in bifundamental representations
$$(1,1,..,{\bf Nd_i},1,...,{\overline {{\bf Nd_j}}},1,..)$$
with multiplicity $n_i^j$ defined above.  For $i=j$ the
above is understood in the sense that we obtain
$n_i^i$  adjoint fields plus $n_i^i$ neutral
fields of $SU(Nd_i)$.
  Note that we can equivalently view
$n_i^j$ as the number of trivial representations in the tensor product

\[
({\bf 4}\otimes R_i\otimes R_j^*)_{trivial}=n_i^j
\]

The asymmetry between $i$ and $j$ is manifest in the above
formula. Thus in general we have
$$n_i^j\not= n_j^i$$ and so the theory in question
is in general a chiral theory.  However
if $\Gamma$ is a real
subgroup of $SU(4)$, i.e. if ${\bf 4}={\bf 4}^*$ as far as
$\Gamma$ representations are concerned,
then we have by taking the complex
conjugate:

\[
n_i^j=({\bf 4} \otimes R_i \otimes R_j^*)_{trivial}=
({\bf 4}\otimes R_i^*\otimes R_j)_{trivial}=n_j^i.
\]

So the theory is chiral if and only if ${\bf 4}$ is a complex
representation of $\Gamma$, i.e. if and only if ${\bf 4}\not={\bf 4}^*$
as a representation of $\Gamma$.
If $\Gamma$ were a real subgroup of $SU(4)$ then
$n_i^j=n_j^i$.

If $\Gamma$ is a complex subgroup, the theory is chiral, but
it is free of gauge anomalies.  To see this, note that
the number of chiral fermions in the
fundamental representation of each group $SU(Nd_i)$ plus $Nd_i$
times the number of chiral fermions in the adjoint representation is given by
\[
\sum_j n_i^j Nd_j=4 Nd_i
\]
(where the number of adjoints is given by $n_i^i$).
Similarly the number of anti-fundamentals plus $Nd_i$ times
the number of adjoints is given by
\[
\sum_j n_j^i Nd_j = \sum Nd_j (4 \otimes R_j \otimes R_i^*)_{trivial} =
\sum Nd_j(4^* \otimes R_j^* \otimes R_i)_{trivial} = 4 Nd_i
\]

Thus we see that
 the difference of the number of chiral fermions
in the fundamental and the anti-fundamental representation
is zero (note that the adjoint representation is real and does
not contribute to anomaly). Thus each gauge group is anomaly free.

In addition to fermions we also have bosons in bi-fundamental
representations.  The number of bosons $M_i^j$in the bi-fundamental representation
of $SU(Nd_i)\otimes SU(Nd_j)$ is given by the number of $R_j$ representations
in the tensor product of the representation ${\bf 6}$ of $SU(4)$
restricted to $\Gamma$ with the $R_i$ representation.  Note that
since ${\bf 6}$ is a real representation we have
$$M_i^j=(6\otimes R_i\otimes R_j^*)_{trivial}=(6\otimes R_i^*\otimes
R_j)_{trivial}=M_j^i$$
In other words for each $M_i^j$ we have a {\it complex} scalar
field in the corresponding bi-fundamental representation.

\bigskip

{\it Interactions.} The interactions of the gauge fields with the matter is fixed by
the gauge coupling constants for each gauge group. The inverse coupling
constant squared for the $i$-th group combined with the theta angle
for the $i$-th gauge group is
$$\tau_i=\theta_i +{i\over 4\pi g_i^2}={d_i \tau \over |\Gamma|}$$
where $\tau=\theta+{i\over 4 \pi g^2}$
is an arbitrary complex parameter independent
of the gauge group and $|\Gamma|$ denotes the number of elements
in $\Gamma$.

There are two other kinds of interactions: Yukawa
interactions and quartic scalar field interactions.
The Yukawa interactions are in 1-1 correspondence with
triangles in the quiver diagram with two directed
fermionic edges and one undirected scalar edge, with compatible directions
of the fermionic edges:
$$S_{Yukawa}={1\over g^2}\sum_{directed \ triangles}
d^{{abc}}{\rm Tr}\psi_{ij^*}^a\phi_{jk^*}^b \psi_{ki^*}^c$$
where $a,b,c$ denote a degeneracy label of the corresponding
fields. $d^{abc}$ are flavor dependent numbers determined by
Clebsch-Gordon coefficients as follows:  $a,b,c$ determine elements $u,v,w$
(the corresponding trivial representation)
in ${\bf 4}\otimes R_i\otimes R_j^*$, ${\bf 6}\otimes R_j\otimes R_k^*$
and ${\bf 4}\otimes R_k\otimes R_i^*$. Then
$$d^{abc}=u\cdot v\cdot w$$
where the product on the right-hand side corresponds
to contracting the corresponding representation indices for $R_m$'s
with $R_m^*$'s as well as contracting the
 $({\bf 4}\otimes {\bf 6}\otimes {\bf 4})$ according to the unique
 $SU(4)$ trivial representation in this tensor product.

 Similarly the quartic scalar interactions
are in 1-1 correspondence with the 4-sided polygons
in the quiver diagram, with each edge corresponding to an undirected
line.  We have
$$S_{Quartic}={1\over g^2}\sum_{4-gons}f^{abcd}\Phi_{ij^*}^a
\Phi_{jk^*}^b\Phi_{kl^*}^c\Phi_{li^*}^d$$
where again the fields correspond to lines $a,b,c,d$ which in turn
determine an element in the tensor products of the form
${\bf 6}\otimes R_m\otimes R_n^*$.  $f^{abcd}$ is obtained
by contraction of the correponding element
as in the case for Yukawa coupling and also using a $[\mu ,\nu][\mu ,\nu]$
contraction in the ${\bf 6}\otimes {\bf 6}\otimes {\bf 6}\otimes {\bf 6}$ part
of the product.

\bigskip

{\it Conformal Theories in 4 Dimensions.} There follows 
a large list of quantum field theories
in 4 dimensions, one for each discrete subgroup of $SU(4)$ and each
choice of integer $N$, motivated from string theory considerations
which has been
proven to have vanishing beta function to leading order in $N$.
Below we argue for the existence of
at least one fixed point even for finite $N$ (under some technical
assumptions).  The vanishing of the beta function at large $N$
can also be argued using AdS/CFT correspondence.

Consider strong-weak duality.  This duality 
exchanges $8 \pi g^2\leftrightarrow
1/8 \pi g^2$ (at $\theta =0$).  This follows from their
embedding in the type IIB string theory which enjoys
the same symmetry.  In fact, this gauge theory {\it defines}
a particular type IIB string theory background
and so this symmetry must be true for the gauge theory
as well.  In the leading order in $N$ the beta function vanishes.
Let us assume at the next order there is a negative beta function,
i.e., that we have an asymptotically free theory.  Then
the flow towards the infrared increases the value of the coupling
constant. Similarly, by the strong-weak duality, the flow towards
the infrared at large values of the coupling constant must
decrease the value of the coupling constant.  Therefore
we conclude that the beta function must have at least one
zero for a finite value of $g$.

This argument is not rigorous for three reasons:  One is that
we ignored the flow for the $\theta $ angle.  This can be
remedied by using the fact that the moduli space is the upper
half-plane modulo $SL(2,Z)$ which gives rise to a sphere
topology and using the fact that any vector field has a zero
on the sphere (``it is impossible to comb the hair on a sphere'').
The second reason is that we assumed asymptotic freedom
at the first non-vanishing order in the large $N$ expansion.
This can in principle be checked by perturbative techniques
and at least it is not a far-fetched assumption.
More serious, however, is the assumption that there is effectively
one coupling constant. It would be interesting to see if one
can relax this assumption, which is valid at large $N$.

\section{Comments on Conformality}

The most exciting aspect of the conformality approach
is in model building beyond the standard model. The reason
the model building is so interesting is that not only the fermion
but also the scalar representations are prescribed by the construction.
Thus one may not simply add whatever Higgs scalars are required 
for the appropriate symmetry breaking. This rigidity is
actually helpful. Lack of adequate space 
precludes including details of the model building described
in\cite{frampton4,frampton5}. Clearly a simple model would encourage
support for this approach. The simplest model using abelian
orbifolds and found in
\cite{frampton5} is based on the gauge group $SU(3)^7$.
This has less generators than the $E(6)$ gauge group and may therefore
be of considerable interest. Non-abelian orbifolds are currently 
under examination.

\bigskip

The final issue concerns gravity. In the CGT for strong and 
electroweak interactions there is no manifest gravity.
One may say there is no evidence for a graviton and 
that one is concerned only with observable physics. Nevertheless 
if one extrapolates to extremely high energy gravity should enter
and it is {\it not} conformally invariant because, of course,
the Newton constant is dimensional.
It would be attractive to understand the incorporation of gravitation
while staying in only
four spacetime dimensions but this possibility remains elusive. 
The CGT itself stands on its own without need of the string
from which its construction was inferred. But to describe gravity
the most promising idea seems to be to add an extra dimension 
and consider $(AdS)_5$. Keeping the full range of the fifth coordinate leads one
back to the absence of gravity on the surface.
But, as pointed out in \cite{verlinde}, truncating the range of the
fifth coordinate leads to a metric field on the 
surface and hence to a graviton.

As a final speculation, is it possible that conformality 
is related to the vanishing cosmological
constant? Until conformal invariance is broken the vacuum energy is zero.
It then depends on how softly conformal invariance can be broken
if a $(TeV)^4$ contribution is to be avoided.
Clearly, the breaking of conformal invariance needs to be studied, not
only for this reason, but also to allow
predictions for dimensionless quantities like mass ratios and mixing angles
in the low-energy theory.

\section{Remarks on Stanley J. Brodsky}

This workshop is in honor of Stan Brodsky's 60th
birthday. 
I first met Stan in 1969 at SLAC when he was 29 years old. 
At that time,
and ever since, as H.C. Pauli said here, he has been very 
welcoming to visitors at SLAC.
Stan is more than a sociable chap since, as R. Braun
informed me, he has published 326 papers
since then, an average of
almost one paper every month. As we have heard today these
papers have had a very significant impact on the development
of quantum electrodynamics
and quantum chromodynamics.
When asked his favorite of his own papers, Stan selected 
an article with Lepage: {\it Exclusive Processes in Perturbative
Quantum Chromodynamics}~~Phys. Rev. {\bf D22,} 2157 (1980).

Stan Brodsky's total work has already attracted well over ten thousand citations.
This puts him in an exclusive club for theoretical physicists and
represents an inspiring career which I hope will continue for decades to come.

\bigskip

\section*{Acknowledgements}

Thanks are due to George Strobel for organizing this 
meeting. This work was supported in part by the US Department of Energy
under Grant No. DE-FG02-97ER41036.

\end{document}